\documentclass[aps,twocolumn,prl,showpacs]{revtex4}
\usepackage{graphicx,amssymb,amsmath}
\usepackage{graphicx}
\usepackage{dcolumn}
\usepackage{bm}

\newcommand{\nt}{$\nu_T=1$}
\newcommand{\dl}{$d/\ell$}
\newcommand{\bpar}{$B_{||}$}
\newcommand{\bperp}{$B_{\perp}$}
\newcommand{\Sxx}{$\sigma_{xx}^{||}$}
\newcommand{\SxxCF}{$\sigma_{xx}^{CF}$}
\newcommand{\iv}{$IV$}

\begin{document}

\title{Exciton Transport and Andreev Reflection in a Bilayer Quantum Hall System}

\author{A.D.K. Finck$^1$, J.P. Eisenstein$^1$, L.N. Pfeiffer$^2$, and K.W. West$^2$}

\affiliation{$^1$Condensed Matter Physics, California Institute of Technology, Pasadena, CA 91125
\\
$^2$Department of Electrical Engineering, Princeton University, Princeton, NJ 08544}

\date{\today}

\begin{abstract} We demonstrate that counterflowing electrical currents can move through the bulk of the excitonic quantized Hall phase found in bilayer two-dimensional electron systems (2DES) even as charged excitations cannot.  These counterflowing currents are transported by neutral excitons which are emitted and absorbed at the inner and outer boundaries of an annular 2DES via Andreev reflection.
\end{abstract}

\maketitle
Bose-Einstein condensation of excitons (electron-hole pairs) was predicted \cite{blatt,moskalenko,keldysh} nearly a half century ago, in the aftermath of the Bardeen-Cooper-Schrieffer theory of superconductivity.  Surprisingly, the first compelling evidence for the phenomenon came from measurements on bilayer two-dimensional electron systems (2DES) in semiconductor heterostructures at high magnetic fields in the quantum Hall effect (QHE) regime \cite{jpemacd}. In analogy with the Cooper pair condensate in a conventional superconductor, an exciton condensate is separated from its charged quasiparticle excitations by an energy gap.  However, for the quantum Hall exciton condensate this is only true in the bulk of the 2D system.  At the edge of the system, where any interface with normal metal contacts must lie, topologically-protected gapless charged excitations are always present.  These gapless excitations complicate the interpretation of transport measurements on bilayer 2DESs supporting an exciton condensate.  Indeed, while much dramatic evidence for exciton transport in such systems has been reported \cite{spielman1,kellogg1,kellogg2,tutuc1,wiersma} it remains essentially indirect and unable to unambiguously demonstrate that excitons are moving through the bulk of the 2D system. In this paper we report just such a demonstration.  

The exciton condensate in bilayer quantum Hall systems occurs at Landau level filling factor $\nu_T =n_T/(eB/h) = 1$, where $n_T$ is the total density of electrons in the bilayer and $eB/h$ is the degeneracy of a single spin-resolved Landau level created by the applied magnetic field $B$.  Numerous experiments have shown that when the interlayer separation $d$ in the bilayer is less than a critical value ($d \lesssim 1.8~\ell$ with $\ell = (\hbar/eB)^{1/2}$ the magnetic length) several dramatic transport anomalies appear at low temperatures.  Prominent among these are a Josephson-like enhancement of interlayer tunneling \cite{spielman1} and the vanishing of the Hall resistance when equal but oppositely directed electrical currents flow through the two layers \cite{kellogg2,tutuc1,wiersma}. The great majority of these previous transport experiments were performed using simply-connected geometries in which all ohmic contacts to the 2DES are on the single outside edge of the sample.  For example, in the counterflow experiments of Kellogg {\it et al.} \cite{kellogg2} which demonstrated the vanishing of the Hall effect, a Hall bar geometry was employed.  These and related experiments \cite{kellogg1,kellogg2,tutuc1,wiersma} found a natural interpretation in the exciton condensation model: counterflowing electrical currents in the two layers might be realized via the unidirectional flow of interlayer excitons.  Being neutral, these excitons experience no Lorentz force and hence exhibit no Hall effect.  However, since all of the electrical contacts resided on the single outside edge of the Hall bar, and are thus unavoidably connected via conducting edge channels harboring charged quasiparticle excitations, these experiments are incapable of directly demonstrating exciton transport through the {\it bulk} of the 2DES.

In order to overcome the intrinsic limitation of Hall bar devices, Tiemann {\it et al.} \cite{tiemann1,tiemann2,tiemann3} explored bilayer transport in an annular Corbino geometry. While conducting edge channels lie along the inner and outer boundaries of the annulus, they are isolated from each other when the \nt\ QHE is well-established and the bulk of the 2D electron system is insulating.  Using this approach Tiemann {\it et al.} \cite{tiemann1,tiemann2,tiemann3} demonstrated a Corbino version of quantized Hall drag \cite{kellogg1}, one of the signature properties of the \nt\ bilayer QHE, and sought to directly observe exciton transport across the bulk of the system by performing the ``loop drag'' experiment suggested by Su and MacDonald \cite{su}. They concluded that an unambiguous identification of exciton transport was not possible in their device, owing to the complicating effects of strong interlayer tunneling \cite{tiemann4}.  This ambiguity is avoided in the present experiments by suppressing tunneling with an in-plane magnetic field and by employing a measurement circuit which directly senses the transport of excitons across the bulk of the 2DES.

\begin{figure}[b]
\begin{center}
\includegraphics[width=1.0 \columnwidth] {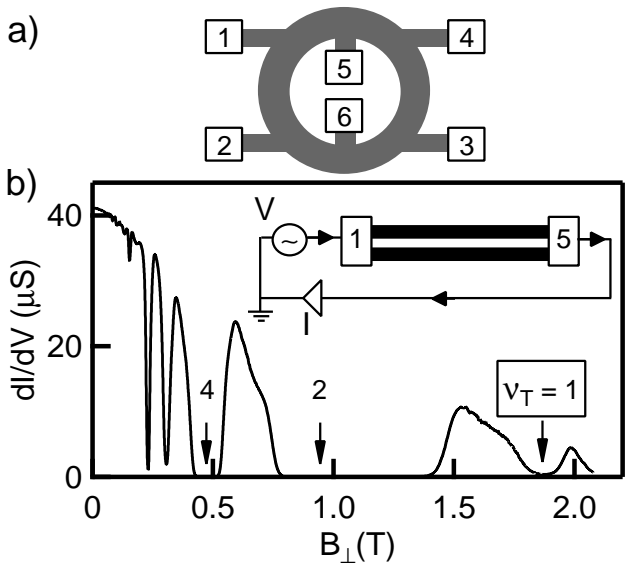}
\end{center}
\caption{a) Schematic of device. b) Corbino conductance vs. \bperp\ at $T = 25$ mK and $\theta = 28^{\circ}$. Inset: Cross-section through annulus, external circuitry, and current-sensitive preamp. Contact 1 is on outer rim, contact 5 on inner rim.}
\label{fig1}
\end{figure}
Our bilayer system is realized in a GaAs/AlGaAs double quantum well (DQW) grown by molecular beam epitaxy.  The DQW consists of two 18 nm GaAs quantum wells separated by a 10 nm Al$_{0.9}$Ga$_{0.1}$As barrier, resulting in a center-to-center layer separation of $d=28$ nm.  Remote Si donors populate each quantum well with a 2DES with an electron density and low temperature mobility of $n_{1,2} \approx 5.5\times10^{10}$ cm$^{-2}$ and $\mu_{1,2} \approx 10^6$ cm$^2$/V-s, respectively.  The sample is patterned into an annulus with inner diameter 1 mm and outer diameter 1.4 mm.  As depicted in Fig.1a, six 100 $\mu$m-wide arms extend from the annulus (two from the inner edge, four from the outer edge) to diffused Ni/AuGe contacts (100x100 $\mu$m$^2$, $\sim 500$ $\Omega$).  Each arm is crossed by narrow front and/or backside metal gates (not shown in the figure). These arm gates allow the ohmic contacts to be switched between four possible states: connected to both layers, to either layer separately, or disconnected from both \cite{sepcon}.  Larger gate electrodes on the top and backside of the sample allow the electron density in each 2D layer within the annulus to be tuned independently; only the balanced $n_1=n_2=n_T/2$ case is considered here. This tunability allows us to reduce the effective layer separation \dl\ to below the critical value of $(d/\ell)_c \approx 1.8$ where interlayer phase coherence and the \nt\ QHE set in.  

At low temperatures and with $d/\ell < 1.8$, our sample clearly exhibits the Josephson-like interlayer tunneling anomaly at \nt\ previously reported \cite{spielman1}. Indeed, the tunneling anomaly is particularly strong here owing to the large area of the Corbino annulus. Since strong tunneling can complicate the interpretation transport studies in the QHE regime (where the sheet conductivity $\sigma_{xx}$ of the 2DES becomes extremely small), our sample is tilted by an angle $\theta$ relative the applied magnetic field $B$.  The resulting in-plane magnetic field component \bpar\ strongly suppresses the Josephson-like tunneling anomaly \cite{spielman2} while not significantly affecting the overall strength of the \nt\ QHE \cite{giudici,finck}.

Figure 1b shows the measured two-terminal Corbino conductance \cite{conductance} \Sxx\ of the sample at $T = 25$ mK and $\theta = 28^{\circ}$ as a function of the perpendicular magnetic field component \bperp.  These data were obtained by applying an ac excitation voltage of 20 $\mu$V at 13 Hz to an ohmic contact on the outside rim of the annulus and detecting the current flowing to ground via an ohmic contact on the inner rim. Since in this case the contacts were connected to both 2D layers simultaneously, the observed conductance reflects parallel current flow in the two layers; this arrangement is illustrated in the inset to the figure.  As expected, deep minima in \Sxx\ are observed whenever the individual 2D layers enter a QHE state ({\it e.g.} at \bperp\ = 0.94 T where $\nu_T = 2 = 1+1$).  In contrast, the minimum in \Sxx\ at \bperp\ = 1.88 T reflects the intrinsically bilayer QHE state at \nt; at this magnetic field neither layer separately would exhibit the QHE.  For Fig. 1b the 2DES densities were adjusted to produce \dl\ = 1.49 at \nt.

We turn now to measurements which rely on contacting the two 2D layers separately.  Figure 2 displays two dc current-voltage (\iv) characteristics taken at \bperp\ = 1.88 T, $T = 25$ mK and $\theta = 28^{\circ}$. Under these conditions \nt\ and \dl\ = 1.49.  In both cases a dc voltage $V$ is applied to one of the ohmic contacts (the source) on the inner rim of the annulus while the current $I$ flowing out of the device to ground via the other ohmic contact (the drain) on the {\it same rim of the annulus} is recorded. The arm gates have been biased to ensure that the source contact is connected only to the top 2DES layer while the drain contact is connected only to the bottom layer.  

For the blue trace in Fig. 2, all four of the ohmic contacts on the outer rim of the annulus have been disconnected from the annulus by appropriately biasing the associated arm gates.  As a result, current flow requires tunneling between the layers.  That very little current is observed to flow attests to the heavy suppression of the coherent Josephson-like tunneling current by the parallel field $B_{||} \approx 1.0$ T.  

\begin{figure}
\begin{center}
\includegraphics[width=1.0 \columnwidth] {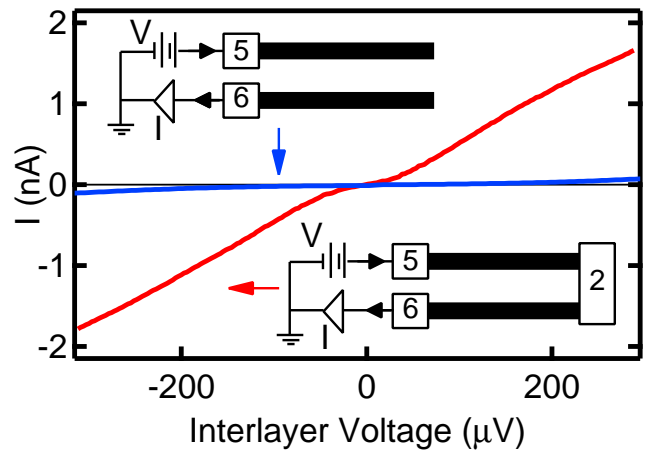}
\end{center}
\caption{Two-terminal \iv\ characteristics at \nt, $T=25$ mK, $B_{\perp}=1.88$ T, and $\theta=28^{\circ}$ with and without the ``short'' on the outer rim. Contacts 5 and 6 are on the inner rim of the annulus; contact 2 is on the outer rim.}
\label{fig2}
\end{figure}
The conditions producing the red trace in Fig. 2 differ from those for the blue trace in only one way: one of the four outer ohmic contacts is re-connected to both 2D layers in the annulus.  Thus this contact ``shorts'' the two layers together at the outside rim.  As Fig. 2 makes clear, this short has a dramatic effect: Far more current flows than in the blue configuration where the short is absent.  

What accounts for the enhanced current in the shorted configuration?  Obviously, if the two layers act independently and have sufficiently large conductivity, current could flow from source to drain by passing first from the inner rim to the outer rim in the top layer, pass through the short, and then return to the drain on the inner rim via the lower layer.  However, the 2DES in the annulus is well within the bilayer \nt\ QHE phase where the layers are very strongly coupled and the conductance \Sxx\ for parallel current flow is quite small, as Fig. 1 demonstrates.  Indeed, if we assume \Sxx\ reflects two independent 2D layers acting in parallel, then we can readily estimate how much current would flow in the configuration corresponding to the red trace.  Even after including the observed dc bias dependence of \Sxx\ \cite{nonlin}, the current so estimated falls more than a factor of 20 below what the red trace in Fig. 2 reveals. 

We believe instead that the enhanced current flow in the shorted configuration is due to exciton transport across the insulating bulk of the 2DES.  A key element of the theoretical descriptions of the coherent \nt\ bilayer is that the conductivity $\sigma_{xx}^{CF}$ for counterflowing (i.e. equal but oppositely directed) currents in the two layers is extremely large even as the conductivity $\sigma_{xx}^{||}$ for parallel currents is exponentially small.  In these theories counterflow transport is completely equivalent to exciton transport. Being neutral, excitons are not confined to the edges of the sample and may move through the bulk.

\begin{figure}
\begin{center}
\includegraphics[width=1.0 \columnwidth] {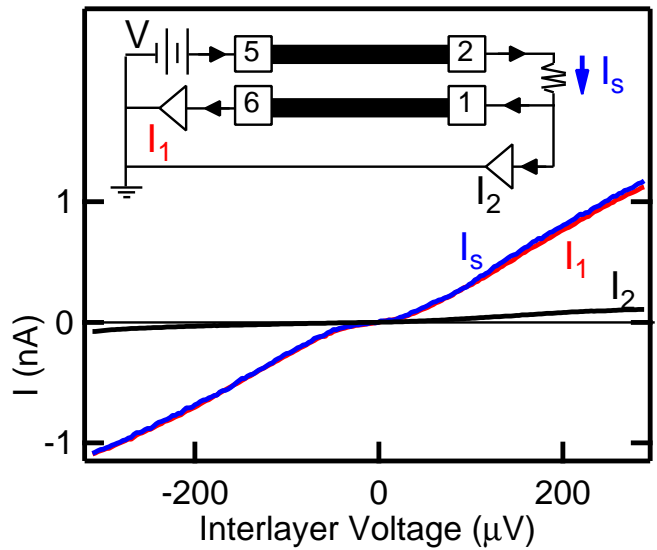}
\end{center}
\caption{\iv\ characteristics at \nt\ with external grounded shunt, at $T = 25$, mK, $B_{\perp}=1.88$ T and $\theta = 28^{\circ}$. Data demonstrate near perfect counterflow $I_s \approx I_1$ and that very little current can be diverted to ground.}
\label{fig3}
\end{figure}
To justify our assertion of exciton transport, we turn to the data in Fig. 3.  Here the short between the two layers at the outside rim of the device is replaced by a 50 k$\Omega$ shunt resistor at room temperature.  This is achieved by connecting the resistor between two of the ohmic contacts on the outide rim, one of which is connected to the top 2D layer and the other to the bottom layer (the two unused ohmics on the outside rim remain disconnected from the annulus).  The voltage drop across this resistor reveals the current flowing between the layers via the shunt.  In addition to grounding the sample at the drain contact on the inner rim as before, a second ground connection is made at one end of the shunt resistor on the outer rim.  A dc voltage is again applied to the top 2D layer at one of the ohmic contacts on the inner rim.  The currents flowing to ground at each of the drains, plus the current flowing through the shunt, are recorded simultaneously.  The inset to Fig. 3 illustrates this set-up \cite{grounds}.

The red trace in Fig. 3 shows the current flowing to ground at the drain on the inner rim of the device; this is analogous to the red trace in Fig. 2.  These data were again acquired at \nt\ and \dl\ = 1.49, with \bperp\ = 1.88 T, $T = 25$ mK and $\theta = 28^{\circ}$.  The blue trace shows the current flowing through the resistive shunt; it is very nearly the same as the red trace. This close equality proves that the enhanced current flow in the shunted configurations of Figs. 2 and 3 does not arise from some new interlayer tunneling process enabled by the shunt.  Instead, this finding reflects counterflowing electrical currents in the two layers crossing the bulk of the 2DES. 

The black trace in Fig. 3 shows that very little current escapes the device via the ground connection at the shunt resistor.  From this we conclude that extremely little {\it net} current flows from the inner to the outer rim of the annulus.  This is of course expected since \Sxx\ is very small due to the QHE.  Nevertheless, the fact that much larger currents flow through the shunt in response to an interlayer voltage applied between the source and drain on the opposite rim of the device, proves that energy is being transported across the insulating bulk of the quantum Hall fluid even as charge is not.

The data in Fig. 3 are consistent with the expectation that the counterflow, or exciton conductivity \SxxCF\ in the bilayer \nt\ QHE is much larger than the parallel flow conductivity \Sxx.  However, we stress that this is an intrinsically bilayer property; one cannot explain the results of Fig. 3 within a picture based upon two independent 2D layers.  Clearly, if the layers were independent, more current would flow to ground at the shunt than would pass back through the high impedance of the lower 2D layer to the ground on the inner rim.  Indeed, the same measurements performed at \nt\ but with $d/\ell =2.3$, where the QHE is absent and the bilayer behaves like two $\nu = 1/2$ composite fermion metals in parallel, show that most of the injected current escapes the device at the grounded shunt. 

\begin{figure}
\begin{center}
\includegraphics[width=1.0 \columnwidth] {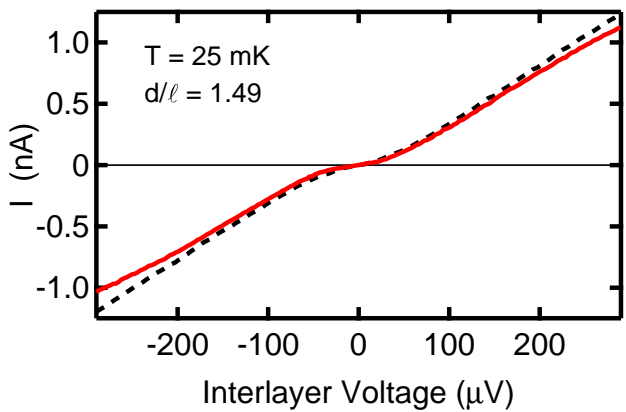}
\end{center}
\caption{Comparison of measured counterflow \iv\ characteristic (solid curve, same as red trace in Fig. 3) at \nt\ with the estimated \iv\ (dashed) due to extrinsic series resistances. Virtually all of the two-terminal voltage drop measured in counterflow can be attributed to the series resistances.}
\label{fig4}
\end{figure}
The above results obtain only at the lowest temperatures $T$ and effective layer separations \dl.  Increasing either increases the current $I_2$ flowing to ground at the shunt and decreases the counterflow current $I_1$.  These are simple consequences increased parallel flow conductance \Sxx\ in the weakened \nt\ QHE state.

Following Su and MacDonald \cite{su} we interpret our results in terms of Andreev reflection.  Electrons entering an edge of the coherent \nt\ region in, say the top 2D layer, are reflected as electrons flowing away from that same edge but in bottom layer.  When tunneling is suppressed this process can only occur if an exciton is launched into the interior of the device \cite{su}.  These neutral excitons may cross the bulk and arrive at the opposite edge.  There the inverse process occurs where electrons are emitted from the top layer and absorbed by the bottom layer; this can only occur if an external current pathway (the shunt) exists.  Within this picture, the emf required to drive current through the shunt is supplied by a time-dependent excitonic order parameter.

It is tempting to interpret the slope of the red \iv\ curves in Figs. 2 and 3 as a measure of the exciton, or counterflow conductance \SxxCF.  Unfortunately this is not possible owing to the two-terminal nature of the measurement.  There are significant extrinsic series resistances, primarily in the arms leading into the annulus, that limit the observed conductance to roughly $3\times 10^{-6}$ $\Omega^{-1}$. We can estimate these resistances via the two-terminal tunneling conductance observed at $\theta = 0$.  At $\theta = 0$ the \nt\ coherent tunneling in our device is so strong that the two-terminal conductance is completely dominated by the extrinsic series resistances.  Separate measurements at the outer and inner rims allow us to construct an composite $IV$ characteristic for the shunted counterflow configuration used in Fig. 3, assuming zero exciton dissipation \cite{dissip}.  The dashed line in Fig. 4 compares this estimate to the actual counterflow conductance (solid trace).  It is clear that the vast majority of the two-terminal voltage drop across our device, as well as the non-linearity near zero bias, results from the series resistances.   

In summary, we have demonstrated that the conductance between contacts on the same edge but opposite layers of the \nt\ bilayer QHE is dramatically affected by a short between the layers on a remote independent edge.  This result, which obtains even when the tunneling between the layers is negligible, demonstrates that charge-neutral excitons propagate through the otherwise insulating bulk of the system.  The emission and absorption of these excitons takes place via Andreev reflection.

We thank A.H. MacDonald for helpful discussions.  This work was supported via NSF grant DMR-1003080.


\begin{references}

\bibitem{blatt} J.M. Blatt, K.W. Boer, W. Brandt, Phys. Rev. {\bf 126}, 1691 (1962).

\bibitem{moskalenko} S.A. Moskalenko, Fiz. Tverd. Tela {\bf 4}, 276 (1962).

\bibitem{keldysh} L.V. Keldysh and Y.V. Kopaev, Fiz. Tverd. Tela {\bf 6}, 2791 (1964).

\bibitem{jpemacd} For a review, see J.P. Eisenstein and A.H. MacDonald, Nature {\bf 432}, 691 (2004) and the references cited therein.

\bibitem{spielman1} I.B. Spielman {\it et al.}, Phys. Rev. Lett. {\bf 84}, 5808 (2000).

\bibitem{kellogg1} M. Kellogg {\it et al.}, Phys. Rev. Lett. {\bf 88}, 126804 (2002).

\bibitem{kellogg2} M. Kellogg {\it et al.}, Phys. Rev. Lett. {\bf 93}, 036801 (2004).

\bibitem{tutuc1} E. Tutuc, M. Shayegan, and D.A. Huse, Phys. Rev. Lett. {\bf 93}, 036802 (2004).

\bibitem{wiersma} R. Wiersma {\it et al.}, Phys. Rev. Lett. {\bf 93}, 266805 (2004).

\bibitem{tiemann1} L. Tiemann {\it et al}, Phys. Rev. B {\bf 77}, 033306 (2008).

\bibitem{tiemann2} L. Tiemann {\it et al.}, New J. of Phys. {\bf 10}, 045018 (2008).

\bibitem{tiemann3} L. Tiemann, Ph.D. thesis, available at http://bieson.ub.uni-bielefeld.de/volltexte/2008/1368/.

\bibitem{su} Jung-Jung Su and A.H. MacDonald, Nature Physics {\bf 4}, 799 (2008).

\bibitem{tiemann4} This is discussed in detail in Ref. \cite{tiemann2}; see section 4.3.  Tunneling was subsequently also found to be strong in the work of Ref. \cite{tiemann1}; see Ref. \cite{tiemann3}, pages 105-106.

\bibitem{sepcon} J.P. Eisenstein, L.N. Pfeiffer, and K.W. West, Appl. Phys. Lett. {\bf 57}, 2324 (1990).

\bibitem{spielman2} I.B. Spielman {\it et al.}, Phys. Rev. Lett. {\bf 87}, 036803 (2001).

\bibitem{giudici} P. Giudici {\it et al.}, Phys. Rev. Lett. {\bf 100}, 106803 (2008).

\bibitem{finck} A.D.K. Finck {\it et al.}, Phys. Rev. Lett. {\bf 104}, 016801 (2010).

\bibitem{conductance} We use \Sxx\ to denote both conductance and conductivity.

\bibitem{nonlin} The \Sxx\ data in Fig. 1 were obtained using a weak, purely ac, voltage excitation.  We find that \Sxx\ gradually increases when a dc bias voltage is added to the ac excitation. This effect is of negligible importance here.

\bibitem{grounds} The currents $I_1$ and $I_2$ flow to ground through current-sensitive preamplifiers with input impedances of 2 k$\Omega$. Replacing the preamp measuring $I_2$ with a low resistance passive ground has no significant impact.

\bibitem{dissip} We assume that a necessary condition for zero exciton dissipation is equal interlayer voltage differences at the inner and outer rims of the annulus. A time-independent order parameter is {\it not} necessary. 

\end{references}
\end{document}